\documentstyle[preprint,aps]{revtex} 

\input{epsf}

\begin{document}



\nopagebreak
\title
{
Dynamical Critical Phenomena and Large Scale Structure of
the Universe: \\
the Power Spectrum for Density Fluctuations. \\
}
\author{J. F. Barbero G.$^{*}$, A. Dominguez$^{*}$, T.  
Goldman$^{+}$ and J.\ P\'erez--Mercader$^{*}$
\vspace{.125in}}
\address{
$^{*}$Laboratorio de Astrof\'{\i}sica Espacial y F\'{\i}sica
Fundamental\\
Apartado 50727\\
28080 Madrid\\
$^{+}$Theoretical Division, Los Alamos National Laboratory\\
Los Alamos, New Mexico 87545\\
\vspace{.125in}
}

\date{July 2nd, 1996}

\maketitle

\vspace{-12.5cm}
\begin{flushright}
LAEFF--96/15\\
July 1996
\end{flushright}
\vspace{9cm}

\begin{abstract}

As is well known, structure formation in the Universe at times
after decoupling can be described by hydrodynamic equations. These  
are shown here to be equivalent to a generalization of the
stochastic Kardar--Parisi--Zhang equation with time--dependent
viscosity in epochs of dissipation. As a consequence of the
Dynamical Critical Scaling induced by noise and fluctuations, these  
equations describe the fractal behavior (with a scale dependent
fractal dimension) observed at the smaller scales for the
galaxy--to--galaxy correlation function and $also$ the  
Harrison--Zel'dovich spectrum at decoupling. By a Renormalization  
Group
calculation of  the two--point correlation function between
galaxies in the presence of ($i$) the expansion of the Universe and  
($ii$) non--equilibrium, we can account, from first principles, for  
 the main features of
the observed shape of the power spectrum.

\end{abstract}

\vspace{.125in}


{\sl Subject headings: Gravitation; Cosmology; Critical Phenomena.}

\vspace{.125in}


Contemporary cosmology is riddled with problems like the Problems  
of the Cosmological Constant, Dark Matter and the Age of the  
Universe. In addition, we have no $dynamical$ understanding of why  
the two--point correlation function for galaxies, a key point of  
contact between theory and observation, has the observed behavior.  
It is known from observations that the
galaxy--to--galaxy correlation function, $\xi_{OBS}(r)$, is well fit by a
power law of the form

$$
\xi_{OBS}(r) \propto r^{-\gamma},
$$

\noindent
where $r$ is the comoving separation between the galaxies and $\gamma$ is
determined from catalogs to be between 1.5 and 1.8 at distances of  
the order of the megaparsec. Nevertheless, at the epoch of  
decoupling, we know from COBE data\cite{gorsky} that the correlation  
function goes like $r^{-(4.2 \pm 0.3)}$. How is this so? How can
this deviation from an integer in the power law exponent be accounted
for? Why is there an evolution in the exponent of the power law?  
Can this fact be established from some generic physics?

In this letter we will $compute$ the galaxy--to--galaxy correlation
function from the hydrodynamics that describes the formation of  
structure in
the Universe and
will present an answer to the above questions.

We generalize previous results
obtained by Berera and Fang in Ref. \cite{fang} and, independently by the
authors of Ref. \cite{gangof four}; we also compute the power  
spectrum of density perturbations. In the first case we have  
generalized their
calculation to include the effects of self--gravity, expansion and,  
simultaneously,
non--equilibrium; we have also generalized their asymptotic
calculation to the full range of distances, from COBE and well into  
the realm of the galaxies. In the latter case, we have included  
deviations from equilibrium and the
expansion of the Universe also in a self--consistent way, and we  
have extended their calculation back into the decoupling era. We  
will not need to introduce any new physics: our
conclusions follow solely from a straightforward (albeit non--naive)
analysis of the hydrodynamic equations and the extension to the Dynamical
Renormalization Group and Dynamical Critical Phenomena of  
techniques familiar from Condensed Matter and Elementary Particle  
Physics.

As argued in Refs. \cite{fang} and \cite{gangof four}, if the power  
law behavior of the two
point correlation function for galaxies is due to some form of
critical phenomena, it follows that in the realm of the galaxies
there must exist some kind of fluctuations which should account
for the observed behavior of $\xi(r)$. We model them by means of  
power law correlated noise.

The suitably averaged value of the
2--point correlation function for the density contrast,  
$\delta(\vec{r},t)$,, written in
comoving coordinates, is identified in phenomenology with
$\xi_{OBS}(r)$ (see, e. g., references \cite{padmanabhan},  
\cite{cole} and \cite{peebles}). We will study the scaling
behavior of the contrast--contrast correlation function.

Under the assumption\footnote{The rotational components of the  
velocity decouple at a quicker rate than their non--rotational  
counterparts, and thus for late enough times it is always possible  
to justify this assumption.} of irrotational peculiar velocity  
$\vec{u}$ and
peculiar acceleration $\vec{w}$, it is straightforward to check that
the hydrodynamic equations arising from the application of Newtonian
considerations to structure formation in the Universe, can be written
in comoving coordinates in terms of a velocity potential $\psi$ and
the gravitational potential $\phi$ due to the contrast as (Ref.  
\cite{thomasetalia}, \cite{fang}, \cite{peebles})

\begin{equation}\label{17}
\frac{\partial}{\partial t} \psi + H \psi - {1 \over 2 a}( \nabla
\psi)^2 - {1 \over a} \phi + {1 \over a} \tilde{f}[\rho_b\{1+(4\pi
Ga^2\rho_b)^{-1} \nabla^2 \phi \}]=0
\end{equation}

\begin{equation}\label{18}
\frac{\partial}{\partial t} \phi + H \phi - 4\pi Ga\rho_b \psi +  
{\cal F}=0
\end{equation}

\begin{equation}\label{19}
\nabla^2 {\cal F} = - \frac{1}{a} \nabla.
[(\nabla\psi)(\nabla^2\phi)].
\end{equation}

\begin{equation}\label{20}
\vec{u}= - \nabla\psi
\end{equation}

\begin{equation}\label{21}
\vec{w}=-a^{-1} \nabla\phi
\end{equation}

\begin{equation}\label{22}
\nabla^2 \phi = 4 \pi G a^2 \rho_b \, \delta (\vec{r},t)
\end{equation}

\begin{equation}\label{22a}
\rho (\vec{r},t) = \rho_b \left [ 1+ \delta (\vec{r},t) \right ] .
\end{equation}

\noindent
Here, $a(t)$ is the scale parameter of the homogeneous cosmological 
background, and the function ${\cal F}(\vec{r},t)$ originates in  
the gauge freedom
associated with the irrotational characters of $\vec{u}$ and  
$\vec{w}$. $H$ is the Hubble parameter. The
function $\tilde{f}$ (also known as minus the specific enthalpy) is  
determined by the equation of state assumed for the
matter whose clustering is described by the above equations: $p=f(\rho)$,
${\tilde f}^\prime(x)=-{f}^\prime(x) / x$. We know however that for  
dust ($p=0$), the Zel'dovich approximation,  $\vec{w}=F(t)\vec{u}$,  
with

\begin{equation}\label{23}
\dot F = 4 \pi G \rho_b - H\,F - F^2
\end{equation}

\noindent
works very well during the early non--linear regime. One can (and  
we will) assume that this approximation is also valid when pressure  
is included and provided that the pressure is small (cf. Ref.   
\cite{thomasetalia}).

Under the above assumptions we have

\begin{equation}\label{25}
 \phi(\vec{x},t) = 4\pi Ga\rho_b  \frac{D(t)}{\dot{D}(t)} \psi(\vec{x},t)
\end{equation}

\begin{equation}\label{25a}
 \delta(\vec{x},t) = \frac{1}{a}  \frac{D(t)}{\dot{D}(t)} \nabla^2  
\psi(\vec{x},t)
\end{equation}

\noindent
where $D(t)$ is the growing mode component of the density
perturbation (Refs. \cite{padmanabhan}, \cite{cole} and  
\cite{peebles}). The velocity potential satisfies

\begin{equation}\label{26}
\frac{\partial}{\partial t} \psi +\frac{\dot{a}}{a} \psi  - {1
\over 2 a}( \nabla \psi)^2 =
4\pi G\rho_b  \frac{D(t)}{\dot{D}(t)} \psi
+\frac{1}{a} c_s^2 \log{\left(1+ \frac{1}{a}
\frac{D(t)}{\dot{D}(t)} \nabla^2 \psi  \right)}\, ,
\end{equation}

\noindent
and expanding the logarithm to lowest order, we get an
equation of the form

\begin{equation}\label{A}
\frac{\partial}{\partial t} \psi = f_1(t) \nabla^2 \psi +f_2(t)(\nabla  
\psi)^2 +f_3(t) \psi \, ,
\end{equation}

\noindent
where the $f_i(t)$ are determined by the background geometry as

\begin{equation}\label{f1}
f_1(t)=\frac{c_s^2}{a^2(t)} \frac{D(t)}{\dot{D}(t)}
\end{equation}

\begin{equation}\label{f2}
f_2(t)=\frac{1}{2a(t)}
\end{equation}

\begin{equation}\label{f3}
f_3(t)=4 \pi G \rho_b(t) \frac{D(t)}{\dot{D}(t)}-\frac{\dot{a}(t)}{a(t)}
\end{equation}

This is a
generalization with time--dependent coefficients (and a mass--like  
term) of the
Kardar--Parisi--Zhang (KPZ) equation

\begin{equation}\label{B}
\frac{\partial}{\partial t}h = \nu \nabla^2 h +\frac{1}{2}\lambda
(\nabla h)^2 +\eta(\vec{x},t) \, ,
\end{equation}

\noindent
which plays a central r\^ole in surface growth phenomena, and whose  
scaling behavior in the IR (large distance, small $k$) and UV  
(short distance, large $k$) regimes are well understood in
terms of the correlation properties of the noise or in the absence  
of noise, the correlation properties of the initial conditions  
(e.g., Ref. \cite{stanley}).

By a series of changes of variables and rescalings, one
can rewrite equation (\ref{A}) as

\begin{equation}\label{C}
\frac{\partial}{\partial \tau } H(\vec{x}, \tau) = \nabla^2  
H(\vec{x}, \tau)
+ (\nabla H(\vec{x}, \tau))^2
+\frac{\partial A(\tau)/\partial \tau}{A(\tau)} H(\vec{x}, \tau)
+\eta(\vec{x}, \tau)
\end{equation}

\noindent
with $A(\tau)=\frac{f_2(\tau)}{f_1(\tau)}\exp{\int^{\tau}_{\tau_0}  
\frac{f_3(\tau')}{f_1(\tau')}d\tau'}$ and the noise term is $\eta =  
(f_2/f_1^2)\bar{\eta}$. The quantity
$H(\vec{x},\tau)$ is related to $\psi(\vec{x},t)$ by
$H(\vec{x},\tau=\int^{\tau} d\tau'  
f_1(\tau'))=\frac{f_2(t)}{f_1(t)}\psi(\vec{x},t)$. The first term in  
the rhs
represents the smoothing effect of diffusion and the second term is due
to non--equilibrium effects. The third term describes the effects  
due to the
time--dependence and matter content of the background geometry (it
contains effects from the expansion of the Universe and  
self--gravity); it behaves,
in the linear approximation and in a variant of ``conformal--time",
as a mass term, and therefore introduces a natural correlation length
into the problem. Finally, the noise term models the various
fluctuations that can appear during epochs in the
evolution of large scale structure in the Universe.

It is clearly seen from Eq. (\ref{C}) that the scaling properties
of $ H(\vec{x}, \tau)$ depend on two key features of the equation: (a)  
the characteristics of the noise and/or (b) the specific features
of the background geometry.

The power spectrum $P(k; t)$ is the Fourier transform of the  
two--point correlation function for the density contrast (see  
\cite{padmanabhan}, \cite{cole} and \cite{peebles}). We  
get\footnote{In what follows and to simplify the writing, we will  
write down $H(\vec{x},\tau)$ instead of $H(\vec{x},\tau=\int^{\tau}  
d\tau' f_1(\tau'))$.}

\begin{equation}\label{pofk}
P(k; t) = \Phi(t) k^2 q^2 {\langle H(\vec{k},\tau)  
\,\,H(\vec{q},\tau) \rangle}_{classical}|_{\vec{q}=-\vec{k}}
\end{equation}

\noindent
where the two--point function for $H(k,\tau)$ does not include the  
effects of non--linearities nor higher order effects due to  
fluctuations, i. e., it is computed from the classical theory, and   
$\Phi^{1/2}(t)=\frac{1}{a(t)}\frac{D(t)}{\partial{D}(t)/\partial  
t}\frac{f_1(t)}{f_2(t)}$ comes from the changes of variables needed  
to transform from Eq. (\ref{22}) to Eq. (\ref{C}).

The renormalized two--point correlation function $\langle  
H(\vec{k},\tau) H(\vec{q}, \tau ') \rangle$ obeys a Callan--Symanzik  
equation, Ref. \cite{freyandtauber}, the solution of which contains  
all its scale dependence, and $P(k;t)$ can be written as

\begin{equation}\label{pofkfull}
P(k; t)_{Full} = \Phi(t) k^2 q^2 {\langle H(\vec{k},\tau)\,\,  
H(\vec{q},\tau) \rangle}_{Improved}
\end{equation}

\noindent
where ${\langle H(\vec{k},\tau) H(\vec{q},\tau)  
\rangle}_{Improved}$ is the solution to the Callan--Symanzik  
equation. As is done within the context of the Renormalization Group  
(RG), this object is computed by inserting into the free two--point  
correlation function\footnote{Obtained from Eq. (\ref{B}) by  
setting $\lambda =0$.} for $\langle H(\vec{k},\tau) H(\vec{q},\tau)  
\rangle$ the values of the couplings obtained by solving their RG  
equations.

We will now restrict ourselves to the following cosmological and  
noise scenario: flat FRW cosmologies where the noise is arbitrarily  
power--law
correlated both in space and in time \footnote{A cosmology with constant
$A$ is flat FRW.}. In flat FRW it is straightforward to see that  
$\Phi_{FRW}(t)=(81/4)c_s^4t_0^{8/3}t^{4/3}$.

For noise with properties given by

\begin{equation}\label{1}
\langle \eta(\vec{k},\omega)\rangle=0
\end{equation}

\begin{equation}
\langle \eta(\vec{k},\omega)\eta(\vec{q},\Omega)\rangle=
2\left[D_0+ D_\theta
k^{-2\rho}\left|\frac{\omega}{\omega_0}\right|^{-2\theta}\right]\delta^3(\vec{k}+
\vec{q})\delta(\omega+\Omega)
\label{2}
\end{equation}

\noindent
a straightforward  calculation gives (here for convergence of some  
integrals, $-1/2 < \theta < 1/2)$

\begin{equation}
\langle H(\vec{k},\tau)\,\,H(\vec{q},\tau)\rangle_{Improved}=
\frac{\nu^2(k)}{2\pi
K_3\lambda^2(k)}\left\{\frac{U_0(k)}
{k^2}+\frac{U_{\theta}(k)\sec(\pi\theta)}{k^{2(1+2\theta+\rho)}}
\right\}\delta^3 (\vec{k}+\vec{q})
\label{31}
\end{equation}

\noindent
where we have introduced $U_0\equiv \frac{D_0\lambda^2}{\nu^3}K_3$,  
$U_\theta\equiv \frac{D_\theta\lambda^2}{\nu^3}K_3$, and $K_3\equiv  
\frac{1}{2\pi^2}$.

The running coupling constants $\nu(k)$, $\lambda(k)$, $U_0(k)$ and  
$U_{\theta}(k)$ obey the following renormalization group equations  
(Ref. \cite{MHKZ}),

\begin{equation}\label{rge1}
-\mu \frac{d\nu}{d\mu}=\nu\left[-\frac{1}{12}U_0+
\frac{2\rho-1}{12}U_{\theta}
(1+2\theta)\sec(\pi\theta)\right]
\end{equation}

\begin{equation}\label{rge2}
-\mu\frac{d\lambda}{d\mu}=\frac{1}{3}\lambda\theta U_\theta(1+2\theta)
\sec(\pi\theta)
\end{equation}

\begin{equation}\label{rge3}
-\mu \frac{d U_\theta}{d\mu}=(4\theta-1+2\rho)U_\theta+
\frac{3+2\theta}{12}U_0
U_\theta+\frac{3+10\theta-6\rho-4\theta\rho}{12}
(1+2\theta)\sec(\pi\theta)U_\theta^2
\end{equation}

\begin{equation}\label{rge4}
-\mu \frac{dU_0}{d\mu}=-U_0+\frac{1}{2}U_{0}^2+\frac{9-6\rho+8\theta}{12}
(1+2\theta)\sec(\pi\theta)U_0 U_\theta+\frac{1}{4}U_\theta^2  
(1+4\theta)\sec(2\pi\theta)\, .
\end{equation}

\begin{figure}
\epsfxsize=15cm 
\epsfbox{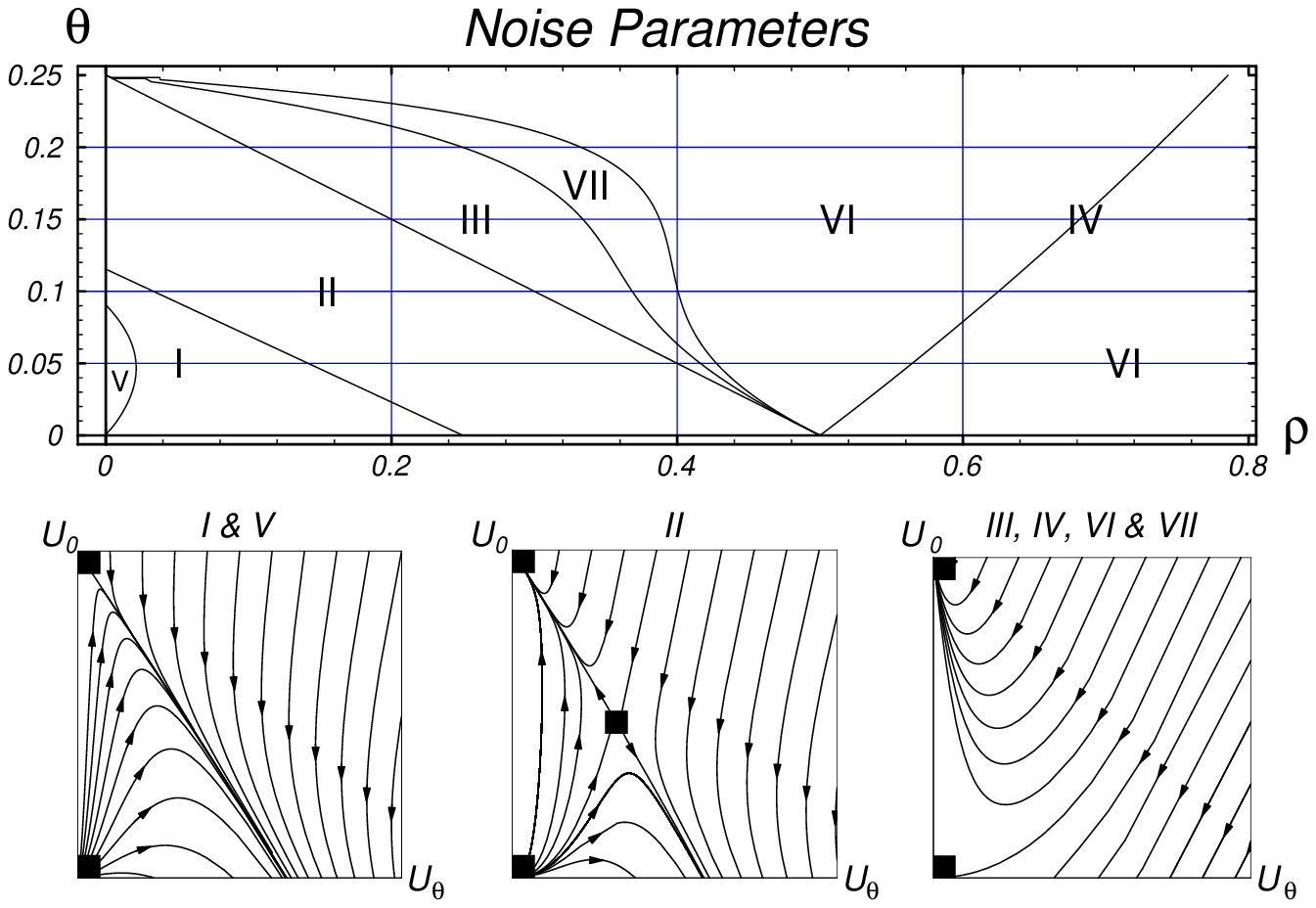}
\caption{Noise spectrum and fixed points for $U_0$ and  
$U_{\theta}$. The fixed point structure for these RG equations  
depends on the noise spectrum. There exist five distinct regions in  
noise parameter space leading to three un--related behaviors of the  
RG equations. In the upper panel we display the noise ``space" and  
in the lower panels the corresponding map of  $U_0$ and $U_{\theta}$  
fixed points together with their IR or UV characters. Different  
choices in the $\rho$--$\theta$ plane lead to qualitatively  
different solutions to the RGEs for $U_0$ and $U_\theta$. In the  
region where correlations are positive, different behaviors can be  
found as shown in this figure. The difference between I and V (or  
the set III, IV, VI and VII) appears in the region where  
$U_{\theta}< 0$ and where the RGEs have different types of critical  
points.
}
\label{fig1}
\end{figure}

\noindent
The parameter $\mu$ has dimensions of momentum. These equations  
have several fixed points, and the asymptotic behavior of the  
solutions depends on the values of the noise parameters $\theta$ and  
$\rho$ (see Fig. \ref{fig1} for a plot and definitions of the  
parameter regions.) The fixed point structure in the region of the  
$(U_{\theta}-U_0)$ plane where the correlations are positive is  
shown in Figure 1.

\begin{figure}
\epsfxsize=14cm 
\epsfbox{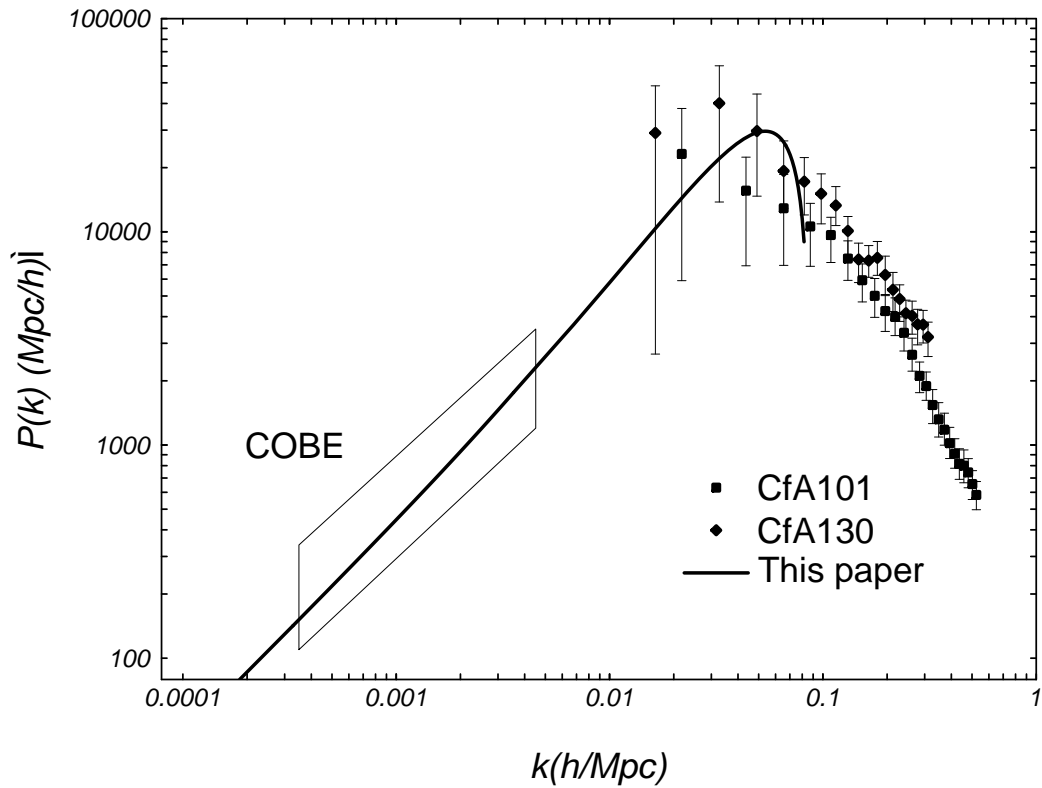}
\caption{The predicted power spectrum for the contrast as a  
function of the momentum scale and some typical observational data.  
The data plotted in the figure are taken from the CfA--101 and  
CfA--130 catalogues (Ref. [10]). The steep fall--off in our  
predicted curve past its maximum indicates the need to include  
higher order terms in the expansion of the logarithm in Equation  
(11) which, as explained in the text, become $relevant$ at shorter  
distances.
}
\label{fig2}
\end{figure}

>From COBE observations we know that in the IR (small $k$--regime)  
the power spectrum is Harrison--Zel'dovich; similarly at ``large"  
momentum, the power spectrum is also scale invariant $but$ with a  
different exponent. Therefore the boundary condition on the improved  
two point function is that for $\vec{k} \rightarrow 0$ the power  
spectrum be Harrison--Zel'dovich, i. e., $\lim_{k \rightarrow 0}  
P_{Full}(k;t) \sim k$. This translates into a condition that must be  
satisfied by the two noise exponents

\begin{equation}
2- 4 \theta - 2 \rho =1\, ;
\end{equation}

\noindent
the subsequent evolution of $P(k)$ is controlled by the RG  
according to the above equations. The results of integrating the RG  
equations in (\ref{rge1}) -- (\ref{rge4}) with a typical set of  
initial conditions compatible with COBE data are shown in Fig.  
\ref{fig2}. They reproduce both the general form of the power  
spectrum and its main observational features. However, as one  
evolves into shorter distances (larger momenta) higher order terms  
in the expansion of the logarithm in the hydrodynamic equation must  
be included since they become $relevant$ at these shorter scales;  
the behavior displayed by the power spectrum in the calculation  
presented here, indicates that the evolution when higher order terms  
are included will also be in the direction of the observational  
evidence, since the largest momenta at which our calculation can be  
trusted is $also$ consistent with the behavior inferred from  
catalogs of galaxies.


We have shown that the hydrodynamics of a fluid of galaxies
interacting through gravity can be studied using scaling techniques  
based on the dynamical renormalization group. We have taken into
consideration all the effects due to self--gravity, expansion of the  
Universe and non--equilibrium present in the ``fluid of galaxies".  
We have applied these ideas to the calculation of the
2--point correlation function for the density contrast, and have
found that (i) its scaling behavior depends on the background
geometry and the noise and/or initial conditions for the density
contrast, but (ii) can be computed and (iii) comparison of our  
results with
observations shows excellent agreement. In summary, we have seen  
that the power spectrum can be viewed as evidence of dynamical  
critical behavior in the Universe. In fact, because of this critical  
behavior, once the power spectrum at decoupling is known to be  
Harrison--Zel'dovich what happens at smaller scales is fairly  
insensitive to small deviations from this initial condition, since  
criticallity implies that the system will eventually be attracted to  
one of its fixed points irrespective of the $details$ of the  
physics.

Although we have demonstrated the feasibility of our approach for  
the simplest background (flat FRW), it is clear that more general  
and complete cases
can be similarly treated. Furthermore, these considerations lead to  
a very interesting view of the large scale structure of the
Universe, where all kinds of new phenomena and behaviors can now be  
described; these include pattern formation, roughening
transitions, nucleation, defect generation, ecological--like  
behavior for galactic and other many body gravitational systems,  
etc..

\acknowledgments{The authors thank Thomas Buchert, Murray
Gell--Mann, Salman Habib, Bernard Jones,  Dennis Sciama, George  
Sewell, Sergei Shandarin and Manuel G. Velarde for useful  
discussions.}

\newpage

\end{document}